\documentclass[letterpaper, 11pt, conference]{ieeeconf}
\usepackage[IEEE]{enderheader}
\newcommand{\Wh}{\hat{W}}
\newcommand{\channel}[1]{\mathsf{#1}}
\newcommand{\Mch}{\channel{M}}
\newcommand{\Wch}{\channel{W}}
\newcommand{\MWch}{\channel{MW}}
\newcommand{\Wmh}{\hat \Wm}

\newcommand{\CM}[1][sum]{C^{{\scriptscriptstyle (\Mch)}}_{#1}}
\newcommand{\CMs}[1][sum]{\tilde{C}^{{\scriptscriptstyle (\Mch)}}_{#1}}
\newcommand{\CW}[1][sum]{C^{{\scriptscriptstyle (\Wch)}}_{#1}}
\newcommand{\CWs}[1][sum]{\tilde{C}^{{\scriptscriptstyle (\Wch)}}_{#1}}

\IEEEoverridecommandlockouts
\begin{document}

\title{\LARGE \bf Achievable Rates for the General Gaussian Multiple Access Wire-Tap Channel with Collective Secrecy
\author{Ender~Tekin and Aylin~Yener%
	\thanks{The authors are with the Wireless Communications and Networking Laboratory,
	Dept. of Electrical Engineering, The Pennsylvania State University,
	University Park, PA 16802 {\tt\small	tekin@psu.edu, yener@ee.psu.edu}}
	\thanks{This work has been supported in part by NSF grant CCF-0514813 ``Multiuser
	Wireless Security"}}}

\maketitle
\thispagestyle{empty}
\pagestyle{empty}

\begin{abstract}
We consider the General Gaussian Multiple Access Wire-Tap Channel (GGMAC-WT).  In this scenario, multiple users communicate with an intended receiver in the presence of an intelligent and informed eavesdropper who is as capable as the intended receiver, but has different channel parameters. We aim to provide perfect secrecy for the transmitters in this multi-access environment. Using Gaussian codebooks, an achievable secrecy region is determined and the power allocation that maximizes the achievable sum-rate is found. Numerical results showing the new rate region are presented.  It is shown that the multiple-access nature of the channel may be utilized to allow users with zero single-user secrecy capacity to be able to transmit in perfect secrecy.  In addition, a new collaborative scheme is shown that may increase the achievable sum-rate.  In this scheme, a user who would not transmit to maximize the sum rate can help another user who (i) has positive secrecy capacity to increase its rate, or (ii) has zero secrecy capacity to achieve a positive secrecy capacity.
\end{abstract}

\normalsize

\section{Introduction}

The wire-tap channel was first analyzed by Wyner in \cite{wyner:wiretap}, where a wire-tapper has access to a degraded version of the intended receiver's signal in a single-user communications scenario.  He measured the amount of ``secrecy" using the conditional entropy of the transmitted message given the received signal at the wire-tapper, and determined the region of all possible Rate/Wiretapper Equivocation pairs.  Wyner showed the existence of a \ital{secrecy capacity}, $C_s$, for communication below which it is possible to transmit zero information to the wire-tapper.  Carleial and Hellman, in \cite{hellman-carleial:wiretap}, showed that it is possible to transmit several low-rate messages at perfect secrecy to achieve an overall rate closer to capacity.  In \cite{leung-hellman:gaussianwiretap}, the authors extended Wyner's results to Gaussian channels and also showed that  Carleial and Hellman's results hold for Gaussian channels as well.  Csisz\'ar and K\"orner, in \cite{csiszar-korner:confbroadcast},
showed that Wyner's results can be extended to weaker, so called ``less noisy" and ``more capable" channels. Furthermore, they analyzed the more general case of sending common information to both the receiver and the wire-tapper.

More recently, the notion of the wire-tap channel was extended to parallel channels, \cite{yamamoto:secretsharing, yamamoto:secretsharinggaussian}, relay channels, \cite{oohama:relaywiretap}, and fading channels, \cite{barros:fadingwiretap}.  Multiple-access channels were considered in \cite{tekin:ASILOMAR05, tekin:ISIT06, liang:genMACconf}.  In \cite{tekin:ASILOMAR05, tekin:ISIT06}, the wire-tapper gets a degraded version of a GMAC uplink signal, and it is shown that the nature of the channel allows an improvement in the individual achievable rates over the single-user channel while having the same limitation on the sum-rate.  In \cite{liang:genMACconf}, there is no external eavesdropper, but the two transmitters try to keep their messages secret from each other.

In \cite{tekin:ASILOMAR05}, we considered the Gaussian Multiple Access Wire-Tap Channel (GMAC-WT) and defined two separate secrecy constraints: (i) the \ital{individual} secrecy constraints, the normalized entropy of any set of messages conditioned on the transmitted codewords of the other users and the received signal at the wire-tapper, and (ii) the \ital{collective} secrecy constraints, the normalized entropy of any set of messages conditioned on the wire-tapper's received signal.  The first set of constraints is more conservative to ensure secrecy of any subset of users even when the remaining users are compromised.  The second set of constraints ensures the collective secrecy of any set of users, utilizing the secrecy of the remaining users.  In \cite{tekin:ASILOMAR05}, we considered a scenario where the wire-tapper received a physically degraded version of the receiver's signal and examined the \ital{perfect secrecy rate regions} for both sets of constraints.  We generalized this to a pre-determined level of secrecy, $0 \le \delta \le 1$, in \cite{tekin:ISIT06, tekin:IT06a}. In this paper, we utilize the collective secrecy constraints with perfect secrecy, and consider the more general case where the eavesdropper's \footnote{Henceforth, we will refer to the adversary as the eavesdropper rather than the wire-tapper since the communication situation modeled is a more general model that is more appropriate for wireless communications} signal is not necessarily degraded, but is at an overall disadvantage compared to the receiver, which we model as a set of received power constraints. Under these constraints, using random Gaussian codebooks, we find an achievable \ital{secure rate region}, where users can communicate with arbitrarily small probability of error with the intended receiver under perfect secrecy from the eavesdropper.  For this achievable rate region, we find the transmit powers that maximize the sum rate.  We also find the sum-rate maximizing power allocation, and users with ``good" channels - those with standardized channel gains below a certain threshold - transmit with maximum power, and those with ``bad" channels, below this threshold, do not transmit.  Next, we show that a non-transmitting user can help increase the secrecy capacity for a transmitting user by effectively ``jamming" the eavesdropper, or even enable secret communications that would not be possible in a single-user scenario.  We term this scheme \ital{collaborative secrecy}.
\newcommand{\lolim}[1]{\text{\b{\ensuremath{#1}}}}
\newcommand{\Csd}[1][\rm{s}]{\ensuremath{\s{C}_{#1}}}
\newcommand{\Gsl}[1][\rm{s}]{\ensuremath{\lolim{\s{G}}_{#1}}}
\newcommand{\Ssc}{\Ss^c}
\newcommand{\DeltaC}{\Delta^{(C)}}
\newcommand{\Pv}{\v{P}}

\section{System Model and Problem Statement}
\label{sec:system}
We consider $K$ users communicating with a receiver in the presence
of an eavesdropper.  Transmitter $k=1,\dotsc,K$ chooses a message $W_k$ from a set of equally likely messages $\Ws_k=\{1, \dotsc, M_k\}$. The messages are encoded using $(2^{nR_k},n)$ codes
into $\{\tilde X_k^n(W_k)\}$, where $R_k=\ninv \log_2 M_k$. 
The encoded messages $\{\tilde   \Xm_k\}=\{\tilde X_k^n\}$ are then transmitted,
and the intended receiver and the eavesdropper each get a copy $\Ym=Y^n$ and
$\Zm=Z^n$.  The receiver decodes $\Ym$ to get an estimate of the transmitted messages, $\Wmh$.  We would like to communicate with the receiver with arbitrarily low probability of
error, while maintaining perfect secrecy, the exact definition of
which will be made precise shortly.

The signals at the intended receiver and the eavesdropper are given by
\newcommand{\hM}{h^{{\scriptscriptstyle (\Mch)}}}
\newcommand{\hW}{h^{{\scriptscriptstyle (\Wch)}}}
\newcommand{\NM}{\Nm^{{\scriptscriptstyle (\Mch)}}}
\newcommand{\NW}{\Nm^{{\scriptscriptstyle (\Wch)}}}
\newcommand{\NMt}{\tilde{\Nm}^{{\scriptscriptstyle (\Mch)}}}
\newcommand{\NWt}{\tilde{\Nm}^{{\scriptscriptstyle (\Wch)}}}
\begin{align}
\Ym &= \ssum_{k=1}^K \sqrt{\hM_k} \tilde \Xm_k + \NMt \\
\Zm &= \ssum_{k=1}^K \sqrt{\hW_k} \tilde \Xm_k + \NWt
\end{align}
where $\NMt,\NWt$ are the AWGN, i.e., $\NMt \isnormal{\v{0},\nvar_\Mch \v{I}}$ and $\NWt \isnormal{\v{0},\nvar_\Wch \v{I}}$.
We also assume the following transmit power constraints:
\begin{equation}
\ninv \sumton{\tilde X_{ki}^2} \le \tilde P_{k,max}, \; k=1,\dotsc,K
\end{equation}

Similar to the scaling transformation to put an interference channel in standard form, \cite{carleial:interference}, we can represent any GMAC-WT by an equivalent standard form as in \cite{tekin:IT06a}:
\begin{subequations}
\label{eqn:YZstd}
\begin{align}
\Ym &= \ssum_{k=1}^K \Xm_k + \NM \\
\Zm &= \ssum_{k=1}^K \sqrt{h_k} \Xm_k + \NW
\end{align}
\end{subequations}
where
\begin{itemize}
\item the original codewords $\{\tilde \Xm\}$ are scaled to get $\Xm_k = \sqrt{\frac{\hM_k}{\nvar_\Mch}}\tilde \Xm_k$.
\item The eavesdropper's new channel gains are given by $h_k = \frac{\hW_k \nvar_\Mch}{\hM_k \nvar_\Wch}$.
\item The noise vectors are normalized such that $\NM = \frac{1}{\nvar_\Mch}\NMt$ and $\NW = \frac{1}{\nvar_\Wch}\NWt$.
\end{itemize}

\begin{figure}[t]
\centering
\includegraphics[height=1.66in,angle=0]{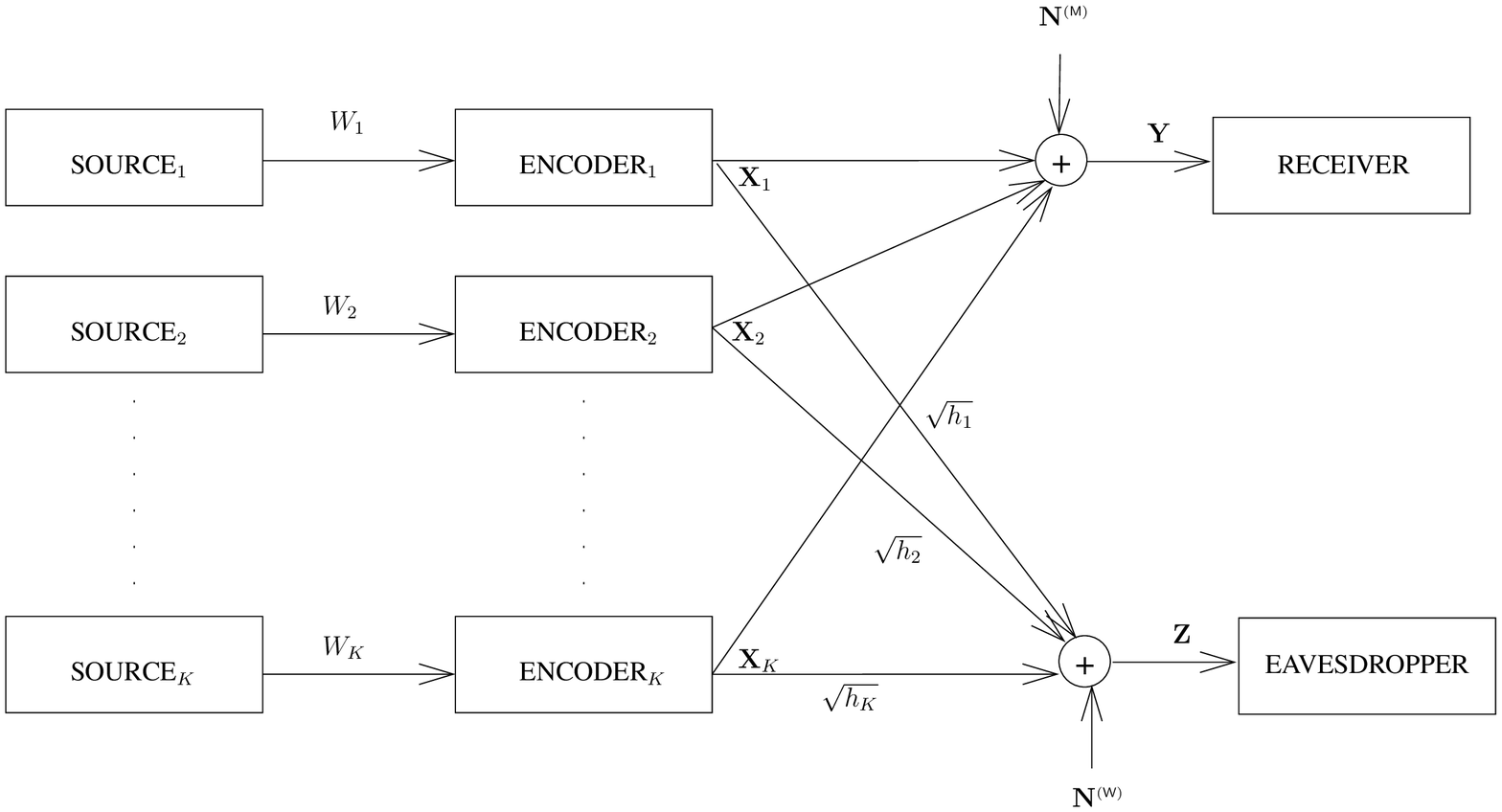}
\caption{\small Equivalent General Gaussian Multiple-Access Wire-Tap Channel (GGMAC-WT) system model.}
\label{fig:gmacwtsystem}
\end{figure}

In \cite{tekin:ASILOMAR05}, we examined the special case of the eavesdropper getting a degraded version of the received signal, which is equivalent to $h_1=h_2=\dotsc=h_K \equiv h < 1$.  In this paper, we look at the more general case where this is not necessarily true. The model is illustrated in Figure \ref{fig:gmacwtsystem}.

We use the collective secrecy constraints defined in \cite{tekin:ASILOMAR05} to take into account the multi-access nature of the channel.
\begin{equation}
\DeltaC_\Ss \triangleq \frac{H(\Wm_\Ss|\Zm)}{H(\Wm_\Ss)} 
	\quad \forall \Ss \subseteq \Ks \triangleq \{1,\dotsc,K\}
\end{equation}

We constrain each subset of users to maintain perfect secrecy, i.e. $\DeltaC_\Ss \ge 1-\e$ for all sets $\Ss$ such that $H(\Wm_\Ss) >0$.  Since this must be true for all sets of users, collectively the system has perfect secrecy.  However, if a group of users are somehow compromised, the remaining users may also be vulnerable.  Note that providing $\DeltaC_\Ks \ge 1-\frac{\e}{r}$, where $r \ge \ssum_{k=1}^K R_k / \min_{k:R_k>0} R_k$ guarantees the perfect secrecy of all subsets as seen by the following argument:
\begin{align}
H(\Wm_\Ks|\Zm) &\ge H(\Wm_\Ks) - \frac{\e}{r} H(\Wm_\Ks) \\
H(\Wm_\Ss|\Zm) &\ge H(\Wm_\Ss)+H(\Wm_{\Ss^c}|\Wm_\Ss) \notag \\
		&\qquad- H(\Wm_{\Ss^c}|\Wm_\Ss,\Zm) -\frac{\e}{r} H(\Wm_\Ks) \\
	&\ge H(\Wm_\Ss) -\frac{\e}{r} H(\Wm_\Ks) \\
\frac{H(\Wm_\Ss|\Zm)}{H(\Wm_\Ss)} &\ge 1 - \frac{H(\Wm_\Ks)}{H(\Wm_\Ss)} \frac{\e}{r} \\
\DeltaC_\Ss &\ge 1-\e 
\end{align}

\begin{definition}[Achievable rates]
\label{def:achrate}
The rate vector $\Rm=(R_1,\dotsc,R_K)$ is said to be \ital{achievable with perfect secrecy} if for any given $\e>0$ there exists a code of sufficient length \n such that
\begin{align}
\hspace{-.1in}
\ninv \log_2 M_k &\ge R_k - \e \quad k=1,\dotsc,K\\
\Perr &\le \e \\
\DeltaC_\Ss &\ge 1-\e \quad \forall \Ss \subseteq \Ks=\{1,\dotsc,K\}
\end{align}
where user $k$ chooses one of $M_k$ symbols to transmit according to the uniform distribution and
\begin{equation}
\Perr = \frac{1}{\prod_{k=1}^K M_k} \hspace{-.15in}
	\sum_{\hspace{.15in}\Wm \in \Ws_1 \times\Ws_2 \times \dotsm \times \Ws_K}
	\hspace{-.3in} \prob{\Wmh \neq \Wm |\Wm \text{ was sent}}.
\end{equation}
is the average probability of error.
We will denote the set of all achievable rates with perfect secrecy $\Csd$.
\end{definition}

Before we state our results, we define the following quantities for any $\Ss \subseteq \Ks$.
\begin{gather*}
\CM[\Ss] \triangleq g\paren{\ssum_{k \in \Ss} P_k}, \hspace{.3in}  
\CW[\Ss] \triangleq g\paren{\ssum_{k \in \Ss} h_k P_k} \\
\CMs[\Ss] \triangleq g\paren{\frac{\ssum_{k \in \Ss} P_k}{1+\ssum_{k \in \Ssc} P_k}} \\
\CWs[\Ss] \triangleq g\paren{\frac{\ssum_{k \in \Ss} h_k P_k}{1+\ssum_{k \in \Ssc} h_k P_k}}
\end{gather*}
where $g(\xi) \triangleq \onehalf \log (1+\xi)$ and $\Ssc=\Ks \setminus \Ss$.  The quantities with $\Ss=\Ks$ will sometimes also be used with the subscript \ital{sum}.  Note that these quantities are functions of $\{P_k\}_{k=1}^K$.  We also define the following set of allowable powers such that $\CM[\Ss] \ge \CWs[\Ss]$ :
\begin{align}
\notag \Ps \triangleq \bigg \{\Pv=(P_1,\dotsc,P_K) \colon& \\
	&\hspace{-.9in} P_{k,max} \ge P_k \ge 0,  && k=1,\dotsc,K, \notag \\
\label{eqn:Pset} &\hspace{-.9in} \phi_\Ss(\Pv) \ge 0 && \forall \Ss \subseteq \Ks \quad \bigg \}
\end{align}
where
\begin{equation}
\label{eqn:phidef}
\phi_\Ss(\Pv) \triangleq \sum_{k \in \Ss} P_k - \frac{\sum_{k \in \Ss} h_k P_k}{1+\sum_{k \in \Ssc} h_k P_k}
\end{equation}

Note that if $h_k \le 1 \, \forall k$, we are left with $\Ps \equiv \{\Pv \colon P_{k,max} \ge P_k \ge 0, \; k=1,\dotsc,K\}$.  On the other hand, if $h_k > 1,\, \forall k$, then the constraint for set $\Ks$ forces $P_k=0, \, \forall k$.
\section{Achievable Rates}
\newcommand{\Xc}{\mathfrak{X}}
\newcommand{\XmS}{\Xm_\Sigma}

Here, we present an achievable region using Gaussian codebooks.  The proof is very similar to the proof of the achievable region presented in \cite{tekin:ISIT06}.  Note that, when $h_1=\dotsc=h_K<1$, the region reduces to the special case examined in \cite{tekin:ASILOMAR05}, \cite{tekin:ISIT06}, \cite{tekin:IT06a}.
\begin{theorem}
\label{thm:achC}
We can transmit with perfect secrecy using Gaussian codebooks at rates satisfying 
\begin{equation}
\label{eqn:Gach}
\ssum_{k \in \Ss} R_k \le \CM[\Ss] - \CWs[\Ss] \quad \forall \Ss \subseteq \Ks
\end{equation}
where $\Pv \in \Ps$. The region containing all $\Rm$ satisfying these equations is denoted \Gsl.
\end{theorem}

\begin{proof}
Let $\Rm=(R_1,\dotsc,R_K)$ satisfy \eqref{eqn:Gach}. For user $k \in \Ks$, consider the scheme:
\begin{enumerate}
\item Let $M_k=\twon{R_k-\e'}$ where $0 \le \e' < \e$ where $\e'$ is chosen to ensure that $M_k$ is an integer.
\item	Generate $2$ codebooks $\Xc_k$ and $\Xc_{kx}$.  $\Xc_k$ consists of $M_k$	codewords, each component of which is drawn $\isnormal{0,\lambda_k P_k -\varepsilon}$. Codebook $\Xc_{kx}$ has $M_{kx}$ codewords with each component randomly drawn $\isnormal{0,(1-\lambda_k) P_k-\varepsilon}$ where $\varepsilon$ is arbitrarily small to ensure that the power constraints on the codewords are satisfied with high probability. Define $R_{kx}=\ninv \log M_{kx}$ and $M_{kt}=M_k M_{kx}$.  Then $R_{kt}=\ninv \log M_{kt}=R_k+R_{kx}+\e'$.
\item To transmit message $W_k \in \{1,\dotsc,M_k\}$, user $k$ finds the codeword corresponding to $W_k$ in $\Xc_k$ and also uniformly chooses a codeword from $\Xc_{kx}$ which are then added and the resulting codeword, $\Xm_k$, is sent so that we are actually transmitting one of $M_{kt}$ codewords.
\end{enumerate}

The specific rates are chosen such that $\forall \Ss \subseteq \Ks$ the following are satisfied:
\begin{align}
\label{eqn:achR} \ssum_{k \in \Ss} R_k &\le \CM[\Ss] - \CWs[\Ss]\\
\label{eqn:achRx} \ssum_{k=1}^K R_{kx} &= \CW \\
\label{eqn:achRt} \ssum_{k \in \Ss} R_{kt} &\le \CM[\Ss]
\end{align}

From \eqref{eqn:achRt} and the GMAC coding theorem, with high probability the receiver can decode the codewords with low probability of error.  We now need to show that the secrecy constraints are satisfied.  Note that since the secrecy of the overall system ensures the secrecy of each subset, we only need to show that the coding scheme described achieves $\Delta_\Ks \ge 1-\e$.  We concern ourselves only with MAC sub-code $\{\Xc_k\}_{k=1}^K$.  From this point of view, the coding scheme described is equivalent to each user $k \in \Ks$ selecting one of $M_k$ messages, and sending a uniformly chosen codeword from among $M_{kx}$ codewords for each.  Define $\XmS=\ssum_{k=1}^K \sqrt{h_k}\Xm_k$.
\begin{align}
\hspace{-.1in} H(\Wm_\Ks|\Zm) \hspace{-.5in} &\\
	&=H(\Wm_\Ks,\Zm)-H(\Zm) \\
	&= H(\Wm_\Ks,\XmS,\Zm)-H(\XmS|\Wm_\Ks,\Zm)-H(\Zm) \\
	&=H(\Wm_\Ks)+H(\Zm|\Wm_\Ks,\XmS)-H(\Zm) \notag \\
		&\hspace{.4in}  +H(\XmS|\Wm_\Ks)-H(\XmS|\Wm_\Ks,\Zm) \\
	&\label{eqn:achprf1}= H(\Wm_\Ks) - I(\XmS;\Zm)+I(\XmS;\Zm|\Wm_\Ks)
\end{align}
where we used $\Markov{\Wm_\Ks}{\XmS}{\Zm} \Rightarrow H(\Zm|\Wm_\Ks,\XmS)=H(\Zm|\XmS)$ to get \eqref{eqn:achprf1}.
We will consider the two terms individually.  First, we have the trivial bound due to channel capacity:
\begin{equation}
\label{eqn:achprf3}
I(\XmS;\Zm) \le n\CW
\end{equation}

Now write
\begin{equation}
I(\XmS;\Zm|\Wm_\Ks) = H(\XmS|\Wm_\Ks)-H(\XmS|\Wm_\Ks,\Zm)
\end{equation}
Since user $k$ sends one of $M_{kx}$ codewords for each message,
\begin{align}
H(\XmS|\Wm_\Ks) &= \log \paren{M_{1x} M_{2x}}\\
	\label{eqn:achprf4a}&= n \paren{R_{1x}+R_{2x}} = n\CW
\end{align}

We can also write
\begin{equation}
\label{eqn:achprf4b}
H(\XmS|\Wm_\Ks,\Zm) \le n\xi_n
\end{equation}
where $\xi_n \tozero$ as $n \toinf$ since, with high probability, the eavesdropper can decode $\XmS$ given $\Wm_\Ks$ due to \eqref{eqn:achRx}.  Note that the individual rates are unimportant - as far as the eavesdropper is concerned, it is receiving one of $n\CW$ codewords with equal probability for each $(W_1,W_2)$ pair.  Using \eqref{eqn:achR}, \eqref{eqn:achRx}, \eqref{eqn:achprf3}, \eqref{eqn:achprf4a} and \eqref{eqn:achprf4b} in \eqref{eqn:achprf1}, we get
\begin{align}
H(\Wm_\Ks|\Zm) 
	&\ge H(\Wm_\Ks)-n\CW+n\CW-n\xi_n\\ 
	\label{eqn:G1}&= H(\Wm_\Ks)-n\xi_n
\end{align}
and dividing both sides by $H(\Wm_\Ks)$ gives

\begin{equation}
\DeltaC_\Ks \ge 1- \frac{\xi_n}{\sum_{k=1}^K R_k}
\end{equation}
completing the proof.  An intuitive way of looking at this is as ``capacity stuffing with superfluous information".  For each message pair, the eavesdropper can decode the extra ``sum-codeword" transmitted if it knew which messages were sent, but since this information arrives at its capacity, it cannot gain any information about the actual transmitted messages.  For a single-user system consisting of user $k$, if $h_k \ge 1$, then the secrecy capacity for that user would have been $0$.  However, the multi-access nature of the channel enables a different user $j$  with $h_j<1$ to ``help" such a user achieve a non-zero rate with perfect secrecy.
\end{proof}

\begin{figure}[t]
\centering
\includegraphics[width=3.0in,angle=0]{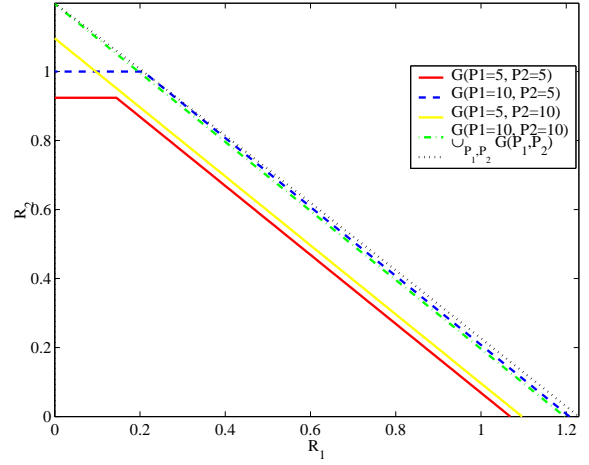}
\caption{\small Achievable rate region for $h_1=.1, \, h_2=.2, \, P_{1,max}=10,\, P_{2,max}=10$}
\label{fig:ggmacwtreg1}
\end{figure}

\begin{figure}[!t]
\centering
\includegraphics[width=3.0in,angle=0]{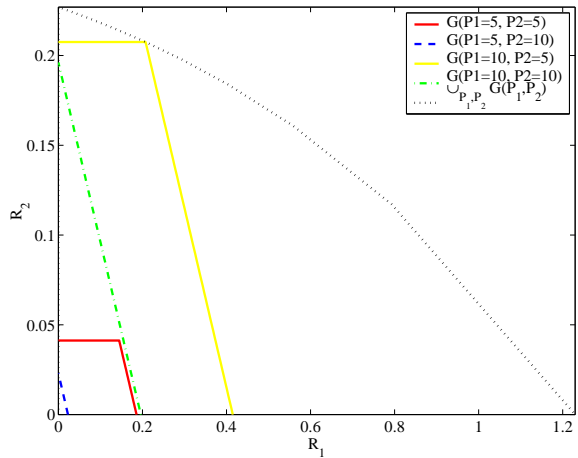}
\caption{\small Achievable rate region for $h_1=.1, \, h_2=1.4, \, P_{1,max}=10,\, P_{2,max}=10$}
\label{fig:ggmacwtreg2}
\end{figure}
\newcommand{\Pvopt}{\Pv^*}
\newcommand{\Pvalt}{\v{Q}}
\newcommand{\Palt}{Q}
\newcommand{\Ts}{\s{T}}
\newcommand{\rhodot}{\dot{\rho}}
\newcommand{\rhodotj}{\rhodot^{(j)}}
\newcommand{\phidotj}{\dot{\phi}^{(j)}}

\section{Maximization of Sum Rate}
The achievable region given in Theorem \ref{thm:achC} depends on the transmit powers.  We are naturally interested in the power allocation $\Pvopt=(\Popt_1,\dotsc,\Popt_K)$ that would maximize the total throughput, i.e. the sum rate.  However, the sum rate maximization is a non-trivial problem since the powers have to be constrained to $\Ps$.  WLOG, assume that the users are ordered such that $h_1 \le h_2 \le \dotsc \le h_K$.
\begin{align}
\max_{\Pv \in \Ps} \; \CM -\CW \hspace{-.6in}& \\
	&= \max_{\Pv \in \Ps} \; g\paren{\sum_{k=1}^K P_k} - g\paren{\sum_{k=1}^K h_k P_k} \\
	&=\min_{\Pv \in \Ps} \; \onehalf \log \rho(\Pv) \\
	&\equiv \min_{\Pv \in \Ps} \; \rho(\Pv) \label{eqn:sumcapprob1}
\end{align}
where we used the monotonicity of the $\log$ function and 
\begin{equation}
\label{eqn:rhodef}
\rho(\Pv) \triangleq \frac{1+\sum_{k=1}^K h_k P_k}{1+\sum_{k=1}^K P_k}
\end{equation}
We start with writing the Lagrangian to be minimized,
\begin{multline}
\label{eqn:Lag}
\hspace{-.12in} \Lag(\Pv,\muv) = \rho(\Pv)
	-\sum_{k=1}^K \mu_{1k}P_k + \sum_{k=1}^K \mu_{2k}(P_k-P_{k,max}) \\
	-\sum_{\Ss \subseteq \Ks} \mu_{3\Ss} \phi_\Ss(\Pv)
\end{multline}
Equating the derivative of the Lagrangian to zero, we get
\begin{multline}
\label{eqn:Lagder}
\frac{\del \Lag(\Pv,\muv)}{\del P_j} = 
	\rhodotj(\Pv) -\mu_{1j} + \mu_{2j} \\
	-\sum_{\Ss \subseteq \Ks} \mu_{3\Ss} \phidotj_\Ss(\Pv)
	= 0
\end{multline}
where
\begin{align}
\label{eqn:rhodotjdef}
\rhodotj(\Pv) &\triangleq \frac{\del \rho(\Pv)}{\del P_j}
	=\frac{h_j - \rho(\Pv)}{1+\sum_{k=1}^K P_k} \\
\label{eqn:phidotjdef}
\phidotj_\Ss(\Pv) &\triangleq \frac{\del \phi_\Ss(\Pv)}{\del P_j} \notag \\
	&=\case
	{1 - \frac{h_j}{1+\sum_{k \in \Ssc} h_k P_k}}{j \in \Ss}
	{\frac{h_j\sum_{k \in \Ss} h_k P_k}{\paren{1+\sum_{k \in \Ssc} h_k P_k}^2}}
		{j \not \in \Ss}
\end{align}
We begin with the following lemma:
\begin{lemma}
\label{lem:zeropow}
Let $\Pvopt$ be the optimum power allocation.  For a user $k \in \Ks$, if $h_k \ge 1$, then $\Popt_k=0$.
\end{lemma}
\begin{proof}
Assume this statement is wrong, i.e., let $\Ts=\{k \in \Ks \colon h_k  \ge 1, \, \Popt_k > 0\} \neq \emptyset$. Consider $\Pvalt$ such that $\Palt_k=\Popt_k, \; k \not \in \Ts$ and $\Palt_k=0, \; k \in \Ts$.  In other words, for a user $k$, $\Palt_k=\Popt_k$ if $h_k <1$ and $\Palt_k=0$ if $h_k \ge 1$.  We first check to see whether $\Pvalt \in \Ps$.  

Since $\Pvopt \in \Ps$, $\Pv_{max} \succeq \Pvalt \succeq \v{0}$, so we only need to check if $\phi_\Ss(\Pvalt) \ge 0 \; \forall \Ss$.
\begin{align}
\phi_\Ss(\Pvalt)
	&=\sum_{j \in \Ss} \Palt_j - \frac{\sum_{j \in \Ss} h_j \Palt_j}
		{1+\sum_{j \in \Ssc} h_j \Palt_j} \\
	&=\sum_{j \in \Ss-\Ts} \Palt_j + \sum_{j \in \Ss\cap\Ts} \Palt_j \notag \\
		&\qquad -\frac{\sum_{j \in \Ss-\Ts} h_j \Palt_j + \sum_{j \in \Ss\cap\Ts} h_j\Palt_j}
		{1+\sum_{j \in \Ssc} h_j \Palt_j} \\
	&=\sum_{j \in \Ss-\Ts} \Popt_j
		-\frac{\sum_{j \in \Ss-\Ts} h_j \Popt_j}{1+\sum_{j \in \Ssc} h_j \Palt_j} \\
	&\ge \sum_{j \in \Ss-\Ts} \Popt_j - \sum_{j \in \Ss-\Ts} h_j \Popt_j \\
	&\ge 0 
\end{align}
since all users $k \in \Ss-\Ts$ must have $h_k<1$.
The proof will be complete if we can show that this new power allocation also increases the sum rate achieved, or equivalently decreases $\rho$.  Begin by writing
\begin{align}
\rho(\Pvalt) 
	&= \frac{1+\sum_{k=1}^K h_k \Palt_k}{1+\sum_{k=1}^K \Palt_k} \\
	&= \frac{1+\sum_{k \in \Ts} h_k \Palt_k+\sum_{k \in \Ts^c} h_k \Palt_k}
		{1+\sum_{k \in \Ts} \Palt_k+\sum_{k \in \Ts^c} \Palt_k}\\
	&= \frac{1+\sum_{k \in \Ts^c} h_k \Popt_k}{1+\sum_{k \in \Ts^c} \Popt_k}\\
	&\le \frac{1+\sum_{k \in \Ts^c} h_k \Popt_k+\sum_{k \in \Ts} h_k \Popt_k}
		{1+\sum_{k \in \Ts^c} \Popt_k+\sum_{k \in \Ts} \Popt_k}\\
	&=\rho(\Pvopt)
\end{align}
where we have used $\frac{a}{b} \le \frac{a+c}{b+d}$ if $\frac{a}{b}\le 1$ and $\frac{c}{d}\ge 1$ when $a,b,c,d \ge 0$.  
\end{proof}
This lemma basically states that to maximize the sum-rate, any user who has a better or equivalent eavesdropper channel must cease transmission.  Now, we look at the optimum power allocation among the remaining users.  This is stated in the below lemma:
\begin{theorem}
\label{thm:sumrate}
The optimum power allocation $\Pvopt$ satisfies $\Popt_k=P_{k,max}$ for $k=1,\dotsc,l$ and $\Popt_k=0$ for $k=l+1,\dotsc,K$ where $l$ is some limiting user such that
\begin{equation}
\label{eqn:sumcaplimusr}
h_l < \frac{1+\sum_{k=1}^l h_k P_{k,max}}{1+\sum_{k=1}^l P_{k,max}} \le h_{l+1}
\end{equation}
\end{theorem}
\begin{proof}
From Lemma \ref{lem:zeropow}, we see that  if $h_k \ge 1$, then $\Popt_k=0$ for all $k \in \Ss$.  Thus, we have $\phi_\Ss(\Pvopt)\ge0$ with equality if and only if $\Popt_k=0$ for all $k \in \Ss$.  Then, from the supplementary conditions, we must have $\mu_{3\Ss}=0$ for all $\Ss$ containing a transmitting user, and $\sum_{k \in \Ss} \Popt_k=0$ for all $\Ss$ not containing a transmitting user.  As a result, $\mu_{3\Ss}\phidotj_\Ss(\Pvopt)=0, \, \forall \Ss \subseteq \Ks$.  Then, it is easy to see that $\Popt_j=P_{j,max}$ if $h_j<\rho(\Pvopt)$ and $\Popt_j=0$ if $h_j > \rho(\Pvopt)$.  A user $j$ may have $0<\Popt_j<P_{j,max}$ iff $h_j=\rho(\Pvopt)$.  However, then the sum rate is independent of that user's power, so we could set $\Popt_j=0$ without any loss in sum rate achievable, and conserve power.
The next step is to find this limiting user $l$.  It is easy to see that this user must satisfy \eqref{eqn:sumcaplimusr}, and can be found in at most $K$ steps.
\end{proof}
\section{Secrecy Through Collaboration}

In the previous section, we showed that the sum secrecy rate is maximized when users with $h_k \ge 1$ do not transmit.  An interesting question in this case, is whether such a user can somehow help increase the secrecy capacity for another user that has $h_k<1$ and is transmitting at full power.  We will show that this is possible in some cases, namely by using the fact that a user with $h_k \ge 1$ can have a `more adverse' effect on the eavesdropper than on the intended receiver.  We will consider the two-user scenario and examine two cases:

\subsection{$h_1 < 1 \le h_2$}
Consider the same case examined in the previous section: $h_1 < 1 \le h_2$.  The sum-rate achievable with perfect secrecy was shown to be $C_s=g(P_{1,max})-g(h_1 P_{1,max})$ with $P_1=P_{1,max}, P_2=0$.  User 2, rather than sit idle, can help user 1 by generating white noise and sending this across the channel.  This will create additional noise at the intended receiver, but even more additional noise at the eavesdropper's receiver.  Since the secrecy capacity in the now single-user channel is known to be the difference of the channel capacities, this scheme may increase the secrecy capacity by reducing the eavesdropper's channel capacity more than it does the intended receiver's.  The problem at hand can be written as:
\begin{multline}
\label{eqn:prob-summaxhelper}
\max_{(P_1,P_2)} g \paren{\frac{P_1}{1+P_2}} - g \paren{\frac{h_1 P_1}{1+h_2 P_2}} \\
	\suchthat \; 0 \le P_1 \le P_{1,max}, \; 0 \le P_2 \le P_{2,max}
\end{multline}
Start by writing the Lagrangian using the monotonicity of $\log$:
\begin{multline}
\label{eqn:Lag-summaxhelper}
\Lag(\Pv,\muv)= - \frac{(1+P_1+P_2)(1+h_2 P_2)}{(1+P_2)(1+h_1 P_1 + h_2 P_2)} \\
	- \sum_{k=1}^2 \mu_{1k}P_k + \sum_{k=1}^2 \mu_{2k}(P_k-P_{k,max})
\end{multline}
Consider user 1:
\begin{multline}
\frac{\del \Lag(\Pv,\muv)}{\del P_1}= \frac{\Psi_1 (P_2)}{(1+P_2)(1+h_1 P_1 + h_2 P_2)^2} \\
	- \mu_{11} + \mu_{21} = 0
\end{multline}
where
\begin{equation}
\hspace{-.01in} \Psi_1 (P_2) = -(1+h_2P_2) \bracket{(1-h_1)+(h_2-h_1)P_2} \hspace{-.3in}
\end{equation}

$\Psi_1(P_2)$ is always negative due to $h_1 < 1 \le h_2$.  Hence, we must have $\mu_{21}>0 \Rightarrow P_1=P_{1,max}$.  Now examine user 2:
\begin{multline}
\frac{\del \Lag(\Pv,\muv)}{\del P_2}= 
	\frac{\Psi_2(P_1,P_2)}{(1+P_2)^2(1+h_1 P_1 + h_2 P_2)^2} \\
	- \mu_{21} + \mu_{22} = 0
\end{multline}
where
\begin{align}
\hspace{-.05in} \Psi_2(P_1,P_2) \hspace{-.4in}&\hspace{.4in}= 
	P_1 h_2 (h_2-h_1) (P_2-p^{(1)})(P_2-p^{(2)}) \\
p^{(1)} & =\frac{-h_2(1-h_1) + \sqrt{D}}{h_2(h_2-h_1)}, \\
p^{(2)} & =\frac{-h_2(1-h_1) - \sqrt{D}}{h_2(h_2-h_1)}, \\
D & =h_1h_2\bracket{(h_2-1)+(h_2-h_1)P_1)}(h_2-1) \hspace{-.2in}
\end{align}
We already know that $P_1=P_{1,max}$.  Note that if $\Psi_2(P_{1,max},P_2) > 0$, then we must have $\mu_{21}>0 \Rightarrow P_2=0$.  On the other hand, if $\Psi_2(P_{1,max},P_2) < 0$, then $\mu_{22}>0 \Rightarrow P_2=P_{2,max}$.  Only when $\Psi_2(P_{1,max},P_2)=0$ do we have $0<P_2<P_{2,max}$.  It is easy to see that $D\ge (h_2-1) \sqrt{h_1h_2} \ge 0$ and hence $p^{(2)} < 0$.  $\Psi_2(P_{1,max},P_2)$ is an upright parabola with respect to $P_2$ with at least one negative root. As a result, if the other root, $p^{(1)}$ is also negative, then $\Psi_2(P_{1,max},P_2)>0 \Rightarrow P_2=0$.  An example is when $h_2=1 \Rightarrow p^{(1)}=p^{(2)}$.  If $p^{(1)}$ is positive, then we have two possibilities, either $P_2$ lies between the roots, in which case $\Psi_2(P_{1,max},P_2) < 0 \Rightarrow P_2=P_{2,max}$, or $P_2 \ge p^{(1)}$.  Since the latter would imply $\Psi_2(P_{1,max},P_2) >0$ and hence $P_2=0$, it is not possible.  Thus, the optimal solution is
\begin{align}
P_1&=P_{1,max}, & \!
P_2&=
	\begin{cases}
	0, & \text{if}\; p^{(1)} \le 0 \\
	p^{(1)}, & \text{if}\; 0< p^{(1)} \le P_{2,max} \\
	P_{2,max}, & \text{if}\; p^{(1)} > P_{2,max} 
	\end{cases}
\end{align}
The condition to have $p^{(1)} \le 0$ is equivalent to $P_{1,max} \le \frac{1-h_1h_2}{h_1(h_2-h_1)}$. Note that if $h_1h_2 \ge 1$, regardless of $P_{1,max}$, user $2$ can always help increase the secrecy capacity.

\begin{figure}[t]
\centering
\includegraphics[width=3.0in,angle=0]{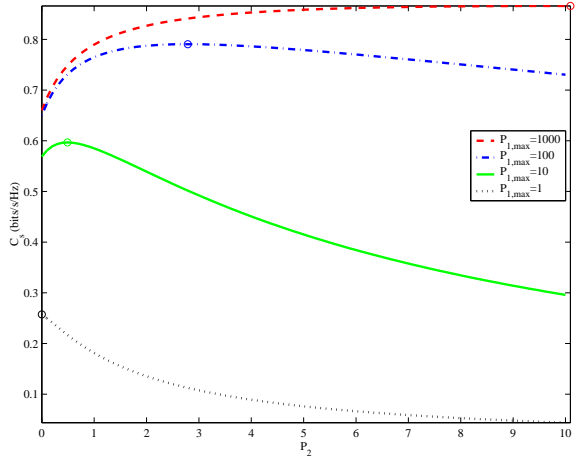}
\caption{\small Sum capacity as a function of $P_2$ with different $P_{1,max}$ for $P_{2,max}=10, \, h_1=.4, \, h_2=1.4$}
\label{fig:ggmacwtcase1}
\end{figure}

\subsection{$1 \le h_1 < h_2$}
Now consider the case where neither user could transmit in the region given in Section 3.  We are motivated by the previous result to see whether user $2$ can actually make it possible for user $1$ to transmit with perfect secrecy.  Our optimization problem and the Lagrangian are the same as given in \eqref{eqn:prob-summaxhelper} and \eqref{eqn:Lag-summaxhelper}.  This time $\Psi_1(P_2)$ has a single root with $P_2 \ge 0$, and is not necessarily negative.  Depending on its value, we will have different optimum $P_1$ values.  Thus,
\renewcommand{\theenumi}{(\roman{enumi})}
\begin{enumerate}
\item $P_2 < \frac{h_1-1}{h_2-h_1} \Rightarrow \Psi_1(P_2) > 0 
	\Rightarrow \mu_{11}>0 \Rightarrow P_1=0$.
\item $P_2 = \frac{h_1-1}{h_2-h_1} \Rightarrow \Psi_1(P_2) = 0 
	\Rightarrow \mu_{11}=\mu_{21}=0$.
\item $P_2 > \frac{h_1-1}{h_2-h_1} \Rightarrow \Psi_1(P_2) < 0 
	\Rightarrow \mu_{21}>0 \Rightarrow P_1=P_{1,max}$.
\end{enumerate}
Now look at user 2.  Again, $D >0$ and $\Psi_2(P_1,P_2)$ is an upright parabola of $P_2$.  However, this time we are guaranteed a positive root as $p^{(1)}>0$, and the solution for $P_2$ depends on $p^{(2)}$.  Consider each of the above cases:  In (i) and (ii), $C_s=0$ regardless of $P_1$, so we are not interested in $P_2$.  Consider case (iii):  We then have $P_2 > p^{(2)}$.  If $P_2< p^{(1)}$, then $\Psi_2(P_1,P_2)<0$, and $P_2=P_{2,max}$.  If $P_2 \ge p^{(1)}$, then $\Psi_2(P_1,P_2) \ge 0$.  Since we cannot have $\mu_{21}>0$ at the same time, the only solution is $P_2=p^{(1)}$.
Summarizing, we get
\begin{equation}
\Pv =
	\begin{cases}
	(0,0), & \text{if}\; P_{2,max} \le \frac{h_1-1}{h_2-h_1} \\
	(P_{1,max},P_{2,max}), & \text{if}\; \frac{h_1-1}{h_2-h_1} < P_{2,max} \le p^{(1)}\\
	(P_{1,max},p^{(1)}), & \text{if}\; P_{2,max} > p^{(1)}	
	\end{cases}
\end{equation}
Note that the solution is of the same form as the previous case.  As long as user $2$ has enough power to make user $1$'s effective channel better than the eavesdropper's, user $1$ can transmit at full power as in the previous setting.  User $1$ could also have helped user $2$, but it is better for the ``worse" user to help the ``better" user to maximize the sum rate.

\begin{figure}[!t]
\centering
\includegraphics[width=3.0in,angle=0]{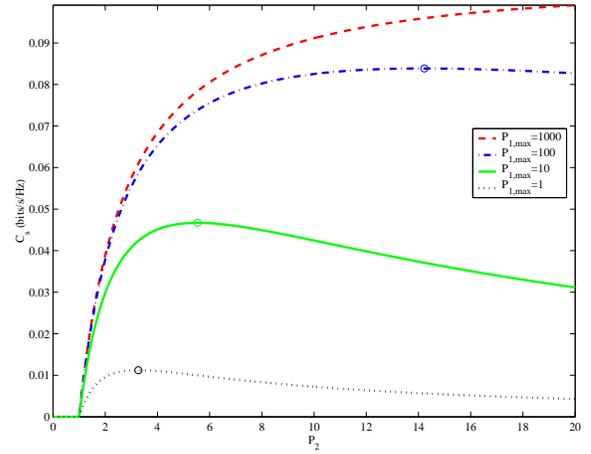}
\caption{\small Sum capacity as a function of $P_2$ with different $P_{1,max}$ for $P_{2,max}=20, \, h_1=1.2,\, h_2=1.4$}
\label{fig:ggmacwtcase2}
\end{figure}
\section{Conclusions}
In this paper, we found an achievable rate region for the General Gaussian Multiple-Access Wire-Tap Channel (GGMAC-WT), in which a second wireless receiver is eavesdropping on the uplink of a GMAC.  We also showed that the sum-rate is maximized when only users with ``better" channels to the intended receiver as opposed to the eavesdropper transmit, and they do so using all their available power.  Moreover we have explored the possibility of the users with worse channels to the intended receiver helping the transmitting users by jamming. This scheme, which we term \ital{collaborative secrecy}, is analyzed for the two user case.

\IEEEtriggeratref{8}
\bibliographystyle{IEEEtran}
\bibliography{IEEEabrv_mod,etekin_full}

\end{document}